\documentclass[a4paper,aps,pra,preprintnumbers,amsmath,amssymb,twocolumn,longbibliography,superscriptaddress]{revtex4-1} 
 \pdfoutput=1 
 
\usepackage{physics}
\usepackage{bbold}
\usepackage{mathtools}
\usepackage[utf8]{inputenc}
\usepackage[T1]{fontenc}
\usepackage{graphicx}
\graphicspath{{figures/}}
\usepackage{amssymb}
\usepackage{amsmath}
\usepackage{algorithmic}
\usepackage{array}

\usepackage{hyperref}
\hypersetup{
    colorlinks=true,
    linkcolor=blue,
    filecolor=magenta,      
    urlcolor=cyan
    }

\usepackage{xcolor} 

\begin{document}

\title{Observing quantum many-body scars in random quantum circuits}

\author{Bárbara Andrade}
\affiliation{ICFO-Institut de Ciencies Fotoniques, The Barcelona Institute of Science and Technology, 08860 Castelldefels (Barcelona), Spain}

\author{Utso Bhattacharya}
\affiliation{ICFO-Institut de Ciencies Fotoniques, The Barcelona Institute of Science and Technology, 08860 Castelldefels (Barcelona), Spain}
\affiliation{Institute for Theoretical Physics, ETH Zurich, 8093 Zurich, Switzerland}

\author{Ravindra W. Chhajlany}
\affiliation{Institute of Spintronics and Quantum Information, Faculty of Physics, Adam Mickiewicz University, 61614 Pozna{\'n}, Poland}

\author{Tobias Gra\ss}
\affiliation{DIPC - Donostia International Physics Center, Paseo Manuel de Lardiz{\'a}bal 4, 20018 San Sebasti{\'a}n, Spain}
\affiliation{IKERBASQUE, Basque Foundation for Science, Plaza Euskadi 5, 48009 Bilbao, Spain}
\affiliation{ICFO-Institut de Ciencies Fotoniques, The Barcelona Institute of Science and Technology, 08860 Castelldefels (Barcelona), Spain}

\author{Maciej Lewenstein}
\affiliation{ICFO-Institut de Ciencies Fotoniques, The Barcelona Institute of Science and Technology, 08860 Castelldefels (Barcelona), Spain}
\affiliation{ICREA, Pg. Lluis Companys 23, 08010 Barcelona, Spain}

\begin{abstract}

The Schwinger model describes quantum electrodynamics in 1+1-dimensions, it is a prototype for quantum chromodynamics, and its lattice version allows for a quantum link model description that can be simulated using modern quantum devices. In this work, we devise quantum simulations to investigate the dynamics of this model in its low dimensional form, where the gauge field degrees of freedom are described by spin 1/2 operators. We apply trotterization to write quantum circuits that effectively generate the evolution under the Schwinger model Hamiltonian. We consider both sequential circuits, with a fixed gate sequence, and randomized ones. Utilizing the correspondence between the Schwinger model and the PXP model, known for its quantum scars, we investigate the presence of quantum scar states in the Schwinger model by identifying states exhibiting extended thermalization times in our circuit evolutions. Our comparison of sequential and randomized circuit dynamics shows that the non-thermal sector of the Hilbert space, including the scars, are more sensitive to randomization.

\end{abstract}

\maketitle

\section{Introduction}

Gauge theories describe fundamental interactions in nature \cite{Burgess2006, Padmanabhan2016}, such as the strong force. Developing quantum devices for quantum simulations of lattice gauge theories is essential as they provide a non-perturbative tool to explore real-time evolution of gauge theories \cite{Banuls2020, Bauer2023, Atas2023}. There are several different quantum platforms for quantum simulators, such as neutral atoms systems \cite{Bluvstein2021, Semeghini2021}, superconducting circuits \cite{Krantz2019, Kjaergaard2020} and trapped ions \cite{Pino2021, Katz2022}. The possibility of engineering multi-body interactions facilitates the quantum simulation of gauge theories. In Ref. \onlinecite{Andrade2022}, a scheme for trapped ions has been proposed which generalizes the Mølmer-Sørensen gate \cite{Molmer1999} and effectively generate Rabi oscillations between the spin states $\ket{\uparrow\uparrow\uparrow}$ and $\ket{\downarrow\downarrow\downarrow}$. The three-body gate $\sigma^+_i\sigma^+_j\sigma^+_k + \rm{h.c.}$ created by this strategy can be used to simulate a U(1) lattice gauge theory. When simulating quantum theories in a quantum computer, we need to express the evolution operator as a product of quantum gates that we understand how to generate. Hence, assuming that the trapped ions strategy is feasible, in this work we trotterize the lattice Schwinger model Hamiltonian to simulate the evolution using these three-body gates.

Our main interest concentrates on the non-equilibrium behavior of this model, especially its thermalization dynamics. Recent advances in quantum technologies allowed the study of many-body systems dynamics out of equilibrium, and the realization of new states of matter. In 2017 \cite{Bernien2017}, a quantum simulator experiment consisting of $51$ Rydberg atoms discovered the existence of initial states with anomalously slow thermalization. The Rydberg atoms system is described by the PXP model, which acts on any sequence of three atoms by exciting/de-exciting the central atom while projecting its neighbors onto the ground state. For this model, the existence of initial states with slow thermalization was explained as a consequence of quantum many-body scars in the spectrum \cite{Turner2018quantum, Turner2018weak}. It has then been recognized that quantum many-body scars \cite{Moudgalya2022} are characteristic for dynamical models with constrained Hilbert spaces \cite{Aramthottil2022, Banerjee2021}, and the study of this phenomenon has become a hot topic in the lattice gauge theory quantum simulation community. Quantum many-body scars are Hamiltonian eigenstates from the middle of the spectrum that highly violate the Eigenstate Thermalization Hypothesis \cite{Deutsch2018, DePalma2015}. Remarkably, in contrast to other mechanisms such as many-body localization and integrability, quantum many-body scars do not depend on the presence of many integrals of motion to exist \cite{Banerjee2021}. In fact, they are observed in interacting, non-integrable and disorder-free models, and have very high overlap with simple many-body states \cite{Chandran2023}. The PXP model can be mapped to a U(1) lattice gauge theory on the line \cite{Surace2020}, the massless Schwinger model. In this study, we investigate how quantum many-body scars manifest in the Schwinger model and study the behavior of states with slow thermalization under Trotterized evolution generated by quantum circuits. We compare using sequential and random quantum circuits to simulate the evolution of states. Random quantum circuits provide a convenient instrument to explore thermalization properties of quantum states \cite{Fisher2023}, and have been used to study measurement-induced phase transitions \cite{Czischek2021}. Moreover, recent quantum computing experiments, such as IBM's $127$-qubit experiment on noise characterization \cite{Kim2023} and Google's $67$-qubit experiment on phase transitions \cite{Morvan2023}, were performed by implementing randomized quantum circuits.

As expected, we find that sequential quantum circuits provide a good approximation to the exact Schwinger Hamiltonian evolution for sufficiently large number of Trotter steps. Also random quantum circuits are able to reproduce the dynamics of the Schwinger model, if both the number  of Trotter steps and circuit realizations is large, yet the agreement  becomes significantly worse for initial state with slow thermalization. This includes the states with large overlap with the scar state Hilbert space sector, but also other states with anomalously slow thermalization. Interestingly, this result indicates that random quantum circuits can be used as a tool to identify states of long thermalization times with information obtained from short time simulations. We note that the Loschmidt echo can also be used to identify these states in simulations of similar time scales.

In the light of the above, this paper is organized as follows: In Section \ref{SectionSchwinger}, we introduce the lattice Schwinger model and rewrite it in a quantum device friendly formulation: The Schwinger quantum link model. In Section \ref{SectionTrotter}, we show how to use quantum circuits with three-body gates to effectively generate the evolution under the full Schwinger quantum link model Hamiltonian, i.e. how to Trotterize the Hamiltonian. In Section \ref{SectionScars}, we discuss the mapping between the PXP model and the Schwinger quantum link model and through correspondence we also highlight the existence of quantum many body scars in both of these models. In Section \ref{SectionResults}, we provide the results obtained by simulating the evolution of states using the quantum circuits.

\section{Lattice Schwinger model} \label{SectionSchwinger}

The Schwinger model is the theory of quantum electrodynamics on the line, i.e. a U(1) gauge theory in $1+1$-dimensions. The Kogut-Susskind lattice formulation of this theory with staggered fermions is described by the Hamiltonian
\begin{align}
    H =& \:J \sum_{j=1}^L \left(\Psi_j^\dagger U_{j,\:j+1} \Psi_{j+1} + \rm{h.c.}\right) + \sum_{j=1}^L E_{j,\:j+1}^2 \nonumber \\
    &+ \mu \sum_{j=1}^L (-1)^j \Psi_j^\dagger \Psi_j,
\end{align}
where the fermionic matter fields $\Psi_j$ and $\Psi_j^\dagger$ live on the lattice sites, particles on the odd sites and anti-particles on the even sites, while the gauge fields $\left(E_{j,\:j+1}, U_{j,\:j+1}\right)$ live on the links connecting lattice sites. The fermionic fields satisfy the anti-commutation relations $\{\Psi_j^\dagger,\Psi_k\}=\delta_{j,\:k}$ and $\{\Psi_j,\Psi_k\}=0$, while the gauge fields satisfy the commutation relation $\comm{E_{j,\:j+1}}{U_{k,\:k+1}}=\delta_{j,k}U_{k,\:k+1}$. We consider $L$ lattice sites, and use periodic boundary conditions $j\sim j+L$ . The three terms in this Hamiltonian are: The kinetic term for the fermion fields, the kinetic term for the gauge fields, and the fermionic mass term. In the lattice formulation, the gauge invariance is encapsulated in the Gauss' law: The difference between the discrete divergence $E_{j,\:j+1}-E_{j-1,\:j}$ and the dynamical charge $\Psi_j^\dagger\Psi_j$ is a conserved quantity. In fact, the operator
\begin{align}
    \tilde{G}_j = E_{j,\:j+1}-E_{j-1,\:j}-\Psi_j^\dagger\Psi_j
\end{align}
commutes with the Hamiltonian, and hence its value is a static charge that defines different sectors of the Hilbert space: $\tilde{G_j}\ket{\Phi_{\rm{phys}}}=q_j\ket{\Phi_{\rm{phys}}}$, where $\ket{\Phi_{\rm{phys}}}$ are the physical states of the theory. Throughout this work, we fix the static charges such that the Gauss' operator takes the form
\begin{align}
    G_j = E_{j,\:j+1}-E_{j-1,\:j}-\Psi_j^\dagger\Psi_j+\frac{1}{2}\left[1-(-1)^j\right]. \label{StaggeredGaussLaw}
\end{align}
This operator annihilates the physical states, i.e. $G_j\ket{\Phi_{\rm{phys}}}=0$.

The Schwinger model admits a quantum link model (QLM) representation, where the gauge fields are described by spin variables: $\left(E_{j,\:j+1}, U_{j,\:j+1}\right)\rightarrow\left(S_{j,\:j+1}^z, S_{j,\:j+1}^+\right)$. After this transformation, we can write the Hamiltonian for the Schwinger QLM. From now on, the gauge fields will carry only one lattice index: They occupy even lattice sites, while fermions occupy the odd ones,
\begin{align}
    H =& \:J \sum_{j\:\rm{odd}} \left(\Psi_j^\dagger\:S_{j+1}^+ \Psi_{j+2} + \rm{h.c.}\right) + \sum_{j\:\rm{even}} \left(S_j^z\right)^2 \nonumber \\
    &+ \mu \sum_{j\:\rm{odd}} (-1)^{(j+1)/2} \:\Psi_j^\dagger \Psi_j\:.
\end{align}
We choose to work on the lowest energy limit of this model, where the gauge fields are spin-$1/2$ variables. The gauge field kinetic term becomes trivial in this limit. We perform a Jordan-Wigner transformation on the fermionic fields, 
\begin{align}
    \Psi_j^\dagger\rightarrow\sigma_j^+\prod\limits_{\substack{k<j\\k\:\rm{odd}}} e^{i\pi(\sigma_k^z+1)/2},
\end{align}
and drop two trivial terms to find the Hamiltonian
\begin{align}
    H =& \: \sum_{j\:\rm{odd}} \left[\frac{J}{2}\left(\sigma_j^+\:\sigma_{j+1}^+ \sigma_{j+2}^- + \rm{h.c.}\right) + \frac{\mu}{2} (-1)^{(j+1)/2} \: \sigma_j^z\right] \:.
\end{align}
Lastly, we perform the unitary rotations $\mathcal{U}=\prod_j\sigma_j^x\sigma_{j+1}^x$, where the product runs only over $j\pmod 4=3$. This rotation transforms the fermionic mass term to a magnetic field proportional to $-\mu/2$. We fix $\mu=0$ and write the energy in units of $J/2$ to reach the form of the Schwinger QLM Hamiltonian used throughout this work,
\begin{align}
    H_{QLM} = \: \sum_{j\:\rm{odd}} \sigma_j^+\sigma_{j+1}^+\sigma_{j+2}^+ +\rm{h.c.}\:. \label{QLMHamiltonian}
\end{align}
We also modify the Gauss' law operator $G_j$ defined in Eq.~\eqref{StaggeredGaussLaw} by using the quantum link model representation, the Jordan-Wigner transformation and unitary rotations $G_j\rightarrow\mathcal{U}^\dagger G_j \mathcal{U}$ to find
\begin{align}
    G_j = \sigma^z_j-\sigma^z_{j-1}-\sigma^z_{j+1}-1. \label{GaussLaw}
\end{align}
The rotations actually give a factor of $-1$ in the Gauss operator for matter sites with $j\pmod 4=1$, but this is not relevant since the physical states are the ones annihilated by this operator. In this formulation, the operator that counts the number of particles and anti-particles is $\sum_{j\:\rm{odd}} (1-\sigma^z_j)/2$.

\section{Trotterized evolution} \label{SectionTrotter}

\begin{figure}
    \centering
    \includegraphics[scale=0.35]{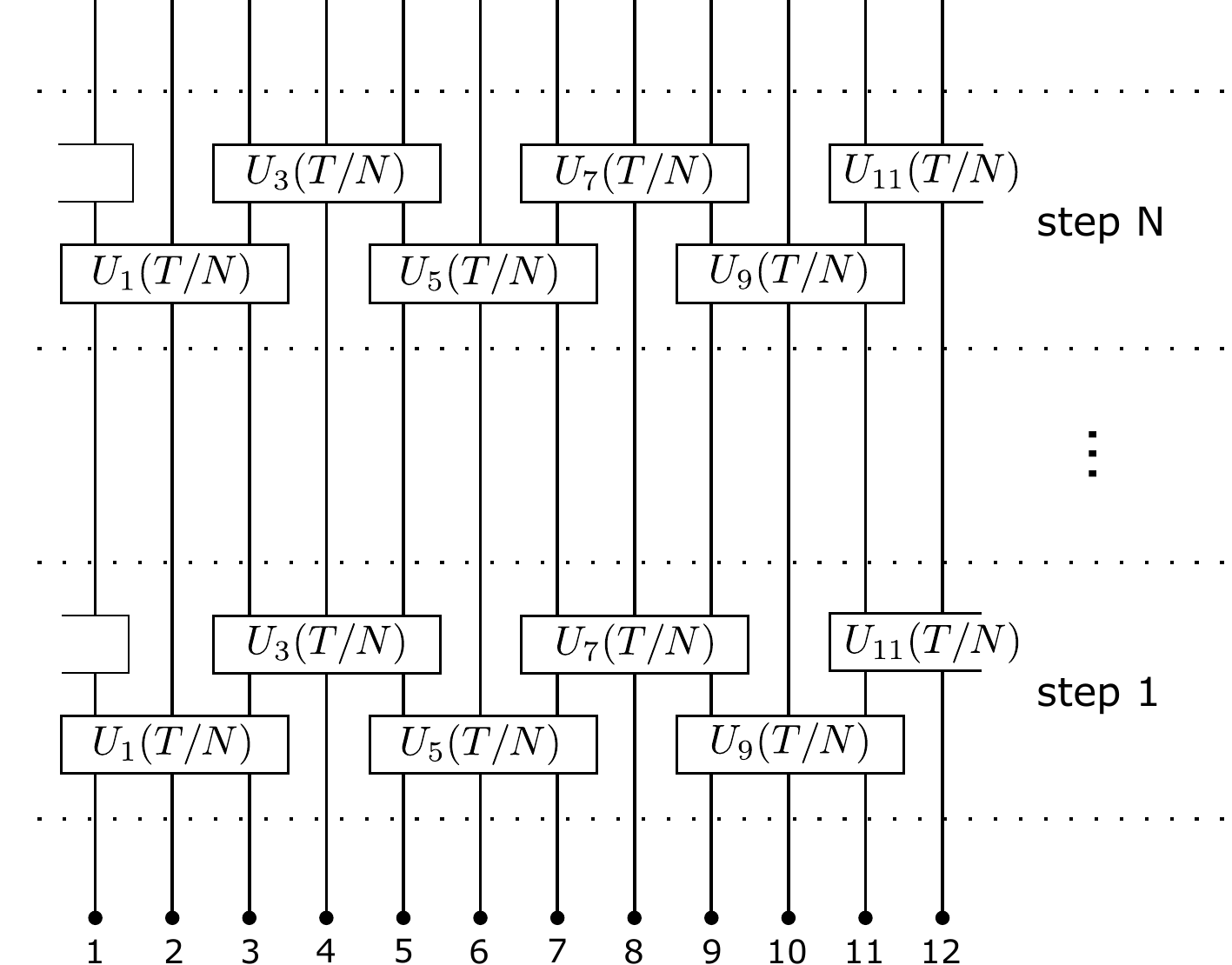} \label{Trotter}
    \caption{Example of quantum circuit corresponding to a sequential trotterized evolution with $12$ qubits. Unitaries are defined as $U_{2j-1}(T/N)\coloneqq e^{-iH_jT/N}$, and $H_j=\sigma^+_{2j-1}\sigma^+_{2j}\sigma^+_{2j+1}+\rm{h.c.}$. This circuit is expected to reproduce the evolution under the full Hamiltonian $e^{-iH_{QLM}T}$ in the limit $N\rightarrow\infty$.}
    \label{circuit}
\end{figure}

In this section, we describe two different quantum circuits that provide approximations to the exact evolution under the Schwinger quantum link model Hamiltonian from Eq.~\eqref{QLMHamiltonian}. The first is a sequential quantum circuit, based on a regular Trotterization of the Hamiltonian, while the second is a random quantum circuit, also based on Trotterization. The Hamiltonian we are interested in simulating has $L/2$ parts, and some of them do not commute, $H_{\rm{QLM}}=\sum_{j=1}^{L/2}H_j$, where 
\begin{equation}
    H_j=\sigma^+_{2j-1}\sigma^+_{2j}\sigma^+_{2j+1} + \rm{h.c.}\:.
\end{equation}
Using the Trotter-Suzuki formula, we approximate the evolution operator $\exp(-iH_{\rm{QLM}}T)$ as a sequence of unitaries $\exp(-iH_j\tau)$ acting for a shorter period of time $\tau=T/N$,
\begin{equation}
    e^{-iH_{\rm{QLM}}T}=\left(\prod_{j=1}^{L/2}e^{-iH_j\tau}\right)^N+\mathcal{O}\left(N\tau^2\right).
\end{equation}
The RHS of this equation defines what we will refer to as the sequential quantum circuit, which consists of $NL/2$ gates. A representation of a sequential circuit of $12$ qubits is provided in Fig.~\ref{circuit}.

The other kind of quantum circuit considered in this work is the random quantum circuit. In this case, we apply $NL/2$ random unitaries $e^{-iH_j\tau}$ for a short period of time $\tau$. Define $\mathcal{R}$ as a set of $NL/2$ random integers $\in[1,L/2]$, the random quantum circuit evolution is defined as 
\begin{equation}
    \tilde{U}_{\cal R}(T) \equiv \prod_{j\in\mathcal{R}}e^{-iH_j\tau}.
\end{equation}
In order to obtain reliable results with random quantum circuits, we average quantities over several runs. For instance, if we want to compute the expectation value of an operator $\hat{O}$ at time $T$, we generate $K$ sets $\mathcal{R}_k$, $k=1,\cdots,K$, each consisting of $NL/2$ random sites, and average out the operator $\hat{O}$ over different random circuit runs. Then the expectation value of $\hat{O}$ at time $T$ is defined as
\begin{equation}
    \expval*{\hat{O}}_{K,T} = \frac{1}{K} \sum_{k=1}^K \bra{\psi_k(T)}\hat{O}\ket{\psi_k(T)},
    \label{statval}
\end{equation}
where $\ket{\psi_k(T)} =  \tilde{U}_{{\cal R}_k}(T)\ket{\psi(0)}$ is the state evolved by the random circuit defined by $\mathcal{R}_k$, and $\ket{\psi(0)}$ is the initial state.

\section{Quantum scars} \label{SectionScars}

\begin{figure*}
    \includegraphics[scale=0.55]{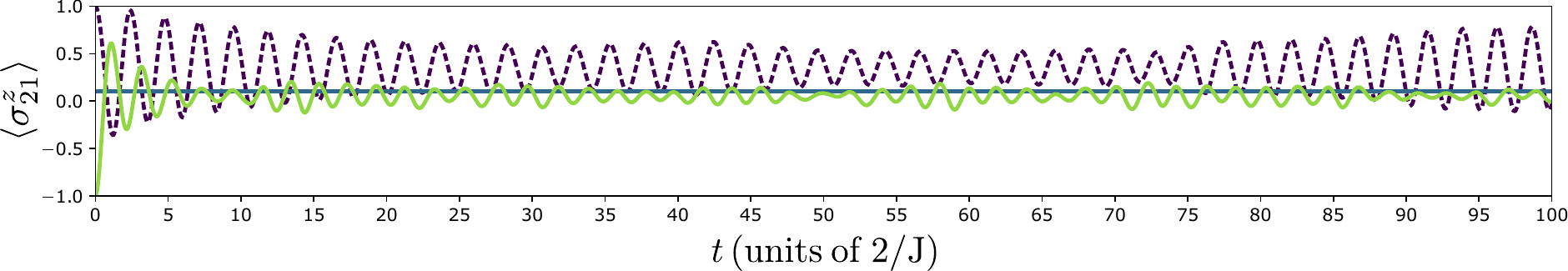}
    \caption{Evolution of expectation value of the local magnetization at the center of the spin chain, at site $j=21$, generated using exact diagonalization. The purple (dashed, dark gray) curve has the vacuum as initial state, while the green (solid, light gray) curve starts at the fully-filled state. The constant curve at $\expval*{\sigma^z_{21}}=0.105$ indicates the thermal value of the local magnetization for both vacuum and fully-filled states.}
    \label{vacuum_fullyfilled}
\end{figure*}

\begin{figure}
    \includegraphics[scale=0.55]{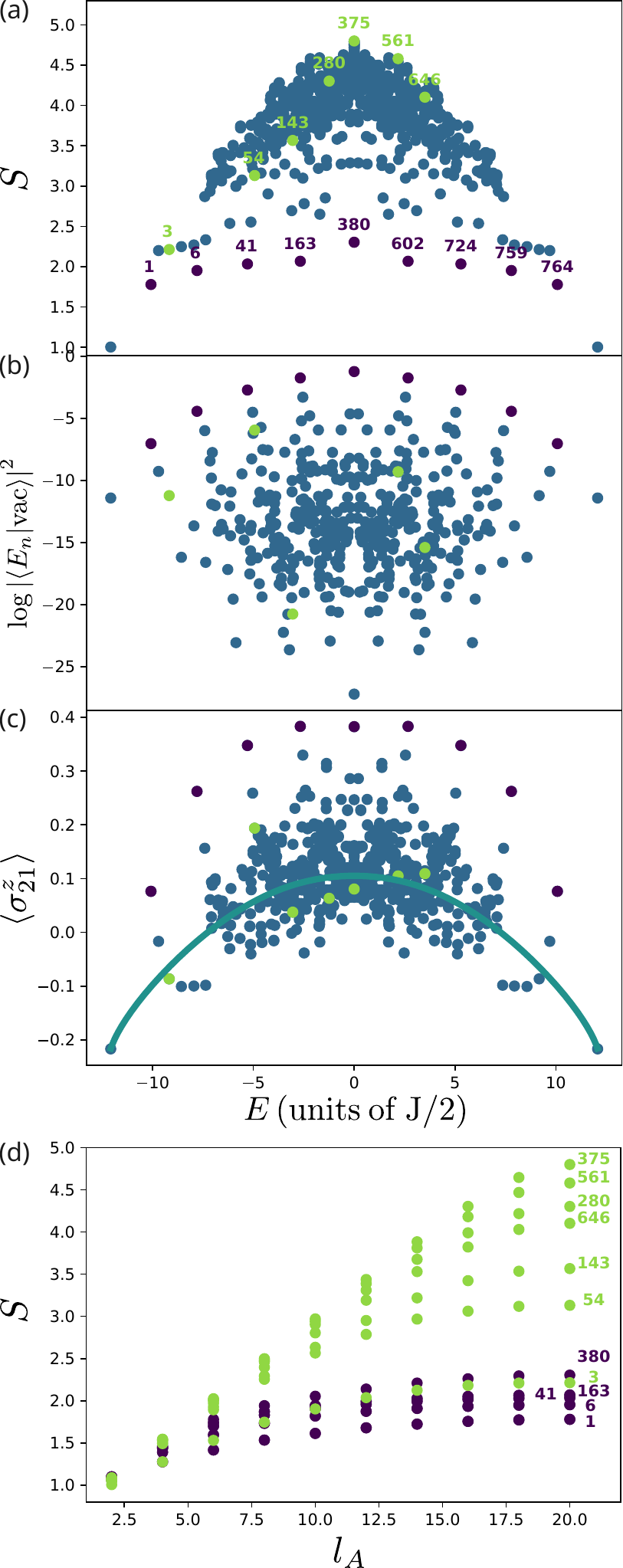}
    \caption{(a) Entanglement entropy of energy eigenstates. Scar states in purple (dark gray), and some typical states, arbitrarily chosen, in green (light gray). (b) Overlap of the energy eigenstates with the vacuum state. Eigenstates with overlap with the vacuum state smaller than $10^{-12}$ are not shown, (c) Expectation value of local magnetization for each energy eigenstate. The curve is the thermal value of the local magnetization. In these plots, some eigenstates $E_n$ are labelled by their energy index $n$, a number that runs from $0$ to $765$, and the scar states are shown in purple (dark gray).}
    \label{scars_qlm}
\end{figure}

In this section, we describe our procedure to find scar states in the Schwinger model. Before that, we introduce the PXP model, which is the model where the quantum many-body scars have been initially studied in the literature. We also describe how to map the PXP model to the Schwinger QLM. The existence of such mapping implies the presence of quantum many-body scars also in the Schwinger QLM, and we identify them in this section.

\paragraph{Scar states in PXP model.} In Ref. \cite{Bernien2017}, the authors realized a 51-qubit quantum simulator consisting of $^{57}$Rb atoms trapped in an optical lattice. The atoms are organized in an one-dimensional array, and they can be either in the ground state $\ket{1}$ or in a highly-excited Rydberg state $\ket{0}$. In this system, nearby excitations are highly suppressed, therefore the atoms dynamics is appropriately captured by the PXP model Hamiltonian
\begin{align}
    H_{\rm{PXP}} = \sum_{j=1}^N P_j \:\sigma_{j+1}^x P_{j+2},
\end{align}
where $P_j$ is the ground state projector, $P_j=\ket{1_j}\bra{1_j}$. The presence of $P_j$ and $P_{j+2}$ enforces the Rydberg blockade mechanism: The evolution under the PXP Hamiltonian does not create nearby excitations.

The authors compared the decay of the domain wall density oscillations for a system of 25 atoms initialized in two different states, the Néel state $\ket{0101\cdots}$ and the ferromagnetic state $\ket{11\cdots}$. They observed that when the initial state is the Néel state, the oscillations last much longer than for the ferromagnetic state. In fact, the Néel state takes longer to thermalize than other "typical" physical configurations.

Following this experiment, a sequence of theoretical studies by C. J. Turner \textit{et al.} \cite{Turner2018weak, Turner2018quantum} indicated that the existence of physical states with long thermalization times is a consequence of highly non-thermal eigenstates in the spectrum, named quantum many-body scars. In the PXP model, the dynamics of physical states is constrained by the Rydberg mechanism, hence the theory weakly breaks ergodicity. As a consequence, in this case we cannot rely on the Eigenstate Thermalization Hypothesis (ETH) to describe the long time behavior of quantum states \cite{Deutsch2018, DePalma2015, Serbyn2021}, i.e. there are states that prevent physical observables from relaxing to the thermal value predicted by the canonical ensemble, even after a long time evolution.

\paragraph{Mapping PXP model onto the Schwinger model.}
In this work, we are interested in states of long thermalization times in the Schwinger model. In Ref. \cite{Surace2020}, the authors describe how to map the PXP model to the Schwinger QLM. The mapping is based on interpreting the spins in PXP model as gauge fields in the Schwinger QLM, in such a way that a system with $N$ spins in the PXP model corresponds to a system with $L=2N$ lattice sites in the Schwinger QLM. More precisely, consider the state $\ket{\alpha_1\alpha_2\cdots\alpha_N}$ in the PXP model, where $\ket{\alpha_j}=\ket{0}$ or $\ket{1}$ at site $j$. We identify $\ket{\alpha_j}$ as the state of the gauge field occupying the lattice site $2j$ in the Schwinger QLM. Then we use the Gauss' law to determine the state of the missing odd lattice sites, which correspond to matter particles. Hence, the mapping between states is $\ket{\alpha_1\alpha_2\cdots\alpha_N} \rightarrow \ket{\beta_1\alpha_1\beta_2\alpha_2\cdots\beta_N\alpha_N}$ with $\ket{\beta_k}$ defined by $G_k\ket{\alpha_{k-1}\beta_k\alpha_k}=0$. On the Hamiltonian level, the mapping between the terms is $P_j\sigma_{j+1}^xP_{j+2}\rightarrow\sigma_k^+\sigma_{k+1}^+\sigma_{k+2}^+ +\rm{h.c.}$, with $k=2j-1$. Creating an excitation in the PXP side corresponds to destroying a pair of particles in the QLM side. For instance, the Néel state $\ket{0101}$ of the PXP model with $4$ sites, which corresponds to the vacuum state $\ket{00010001}$ in the Schwinger QLM, and the ferromagnetic state $\ket{1111}$ corresponds to the fully-filled state $\ket{11111111}$.
With the mapping between the PXP model and the Schwinger QLM established,  scar states must occur in the Schwinger QLM side as well. 

\paragraph{Scars in the Schwinger model.} We use three quantities to identify which eigenstates are the scar states of the Schwinger model: the overlap with the vacuum state, the scaling of the entanglement entropy, and the dynamics of the local magnetization, which we elaborate below. We work with the Schwinger QLM with $L=40$ sites, and use exact diagonalization methods in this subsection.

The Schwinger QLM Hamiltonian is symmetric under translations of two sites, hence we can use the eigenstates of the two-site translation operator $T_2$ to diagonalize the Hamiltonian $H_{QLM}$. The action of $T_2$ in a physical \footnote{by physical we mean that $\ket{f}$ is a spin configuration allowed by the Gauss' law: $G_j\ket{f}$=0 for all odd $j$.} configuration $\ket{f}$ generates a cyclic family of configurations connected by $T_2$, with multiplicity $m_f$. For instance, the vacuum configurations $\ket{f}=\ket{0001}$ and $T_2\ket{f}=\ket{0100}$ are a family of multiplicity $2$. We make superpositions of the states in the same family to build the eigenstates of $T_2$,
\begin{align}
    \ket{\Phi_{p,f}}=\sum_{k=1}^{m_f}\frac{e^{ipk}}{\sqrt{m_f}}T_2^{k-1}\ket{f}, \label{eqclass}
\end{align}
with eigenvalue $e^{-ip}$, where the momentum can assume values $p=2\pi n/m_f,\:n\in\mathbb{Z}$. For $L=40$ lattice sites, the Hilbert space has dimension $766$ \footnote{The physical Hilbert space dimension, before taking translation symmetry into consideration, is $\rm{dim}(\mathcal{H}^{1/2}_L) = 2 + \sum_{i=1}^{\frac{\frac{L}{2} - 2}{2}} \frac{\left(2i + \frac{\frac{L}{2} - 2i}{2}\right)!}{(2i)! \left(\frac{\frac{L}{2} - 2i}{2}\right)!} + 1 +  \sum_{i=1}^{\frac{\frac{L}{2} - 4}{2}} \frac{\left(2i + \frac{\frac{L}{2} - 2 - 2i}{2}\right)!}{(2i)! \left(\frac{\frac{L}{2} - 2 - 2i}{2}\right)!} + 1$. For gauge fields of spin $s$, the physical Hilbert space dimension generalizes to $\rm{dim}(\mathcal{H}^s_L) = \rm{dim}(s)+(\rm{dim}(s)-1)(\rm{dim}(\mathcal{H}^{1/2}_L)-2)$}. Throughout this project, we work on the zero-momentum sector, and we work with two basis sets: The physical basis $\{\ket{\Phi_{f}}\}$ of $T_2$ eigenstates, where $p=0$ is implied, and the energy eigenstates $\{\ket{E_n}\}$.

As discussed before in Ref.\cite{Bernien2017}, the Néel state has an anomalous slow thermalization time when compared to a typical configuration such as the ferromagnetic state. In the Schwinger QLM, the Néel state corresponds to the vacuum, while the ferromagnetic state corresponds to the fully-filled state. In Fig.~\ref{vacuum_fullyfilled}, we compare the evolution of a local observable, the magnetization at the center of the spin chain, for these two states. At $t=10$, the fully-filled state is already thermalized, since the local magnetization after this point only oscillates with low amplitude around the thermal value \footnote{Given a state $\ket{\psi}$, we fix its inverse temperature $\beta$ using the equation $\Tr(\rho_\psi H) = \bra{\psi}H\ket{\psi}$, where $\rho_\psi$ is the thermal state $\rho_\psi=e^{-\beta H}/Z$, and $Z=\Tr(e^{-\beta H})$. Then we use $\rho_\psi$ to compute the expectation value of the operator $\hat{O}$ as $\Tr (\rho_\psi \hat{O})$.} of the local magnetization. On the other hand, for the vacuum state even longer after $t=10$, the local magnetization still oscillates with high amplitude. For this reason, in the Schwinger QLM we use the overlap with the vacuum state to identify scar eigenstates. 

In Fig.~\ref{scars_qlm}(a), we have the entanglement entropy as a function of the energy for all eigenstates $\ket{E_n}$. The Hilbert space partition used to compute the entanglement entropy is $\mathcal{H}_A\otimes\mathcal{H}_B$, where $\mathcal{H}_A$ is the first $L/2$ spins and $\mathcal{H}_B$ is the other $L/2$ spins. For each eigenstate, we compute the entanglement entropy $S(\rho_A)=-\Tr_A(\rho_A\log\rho_A)$, where $\rho_A=\Tr_B(\rho_n)$ and $\rho_n=\ket{E_n}\bra{E_n}$. The majority of eigenstates have entanglement entropy $S>3$. In Fig.~\ref{scars_qlm}(d), we show how the entanglement entropy scales for some states, including those with low entanglement entropy. Most eigenstates are distributed as expected by ETH following a volume law. There are, however, a class of states with sub-volume behavior, which violate ETH. In Fig.~\ref{scars_qlm}(b), we have the overlap between the energy eigenstates and the vacuum state. As discussed in the previous section, the vacuum state in the Schwinger QLM corresponds to the Néel state of the PXP model, and both have much longer thermalization times than a typical physical configuration such as the fully-filled state. The vacuum state also has several revivals in the Loschmidt echo, defined as the overlap between the evolved state and the initial state, which is another indication of slow thermalization. Hence, we use the overlap as a signature to track which eigenstates are behind long thermalization behavior of the vacuum. In Fig.~\ref{scars_qlm}(c), we take a local operator, the magnetization in the center of the spin chain, and compute its expectation value for each eigenstate. We compare this with the thermal value expected for this quantity, and we observe that some states highly deviate from the thermal prediction. The points highlighted in purple (dark gray) are the same eigenstates through Figs.~\ref{scars_qlm}(a)-(c), and these have all of the following properties: low entanglement entropy, high overlap with the vacuum, and they highly violate the thermal expectation of a local observable. For these reasons, we will refer to them as the \emph{quantum many-body scar states} of the Schwinger QLM for $L=40$ lattice sites.

In our investigation, we notice that a superposition of eigenstates equally spaced in energy will always have revivals in the Loschmidt echo. Consider the initial state $\ket{\psi_0}=\sum_{n=0}^N \left(\ket{-n\omega} + \ket{n\omega}\right) / \sqrt{2N}$, where $H_{\rm{QLM}}\ket{n\omega} = n\omega \ket{n\omega}$. The Loschmidt echo at time $t$ is $\left| \sum_{n=0}^N \cos(n\omega t)/N \right|^2$, and this function reaches unity at times $t=2\pi k/\omega$ for $k=0,1,2,\dots$. Thus, the occurrence of revivals in the Loschmidt echo is expected when the superposed eigenstates are evenly spaced in energy. In both the Schwinger QLM and the PXP model, the scar states are nearly equidistant in energy, leading to revivals in the Loschmidt echo when superposed. However, a state exhibiting revivals in the Loschmidt echo may not necessarily strongly overlap with the scar states; rather, it may belong to another class of non-thermal states. We also notice that the Schwinger QLM may have other classes of scar states. In the entanglement entropy scaling plot, Fig.~\ref{scars_qlm}(d), we see that the scaling of eigenstate $\ket{E_3}$ is closer to the scars than to the other typical eigenstates. In Figs.~\ref{scars_qlm}(b) and (c), we see that there is an arc of states right below the quantum many-body scars in purple (dark gray), that also highly overlap with the vacuum and violate the thermal expectation for the local magnetization. Our numerical simulations show that a superposition of these states also have slow thermalization times for the local magnetization, and for these reasons we believe they could be regarded as a different class of quantum scar states.

\section{Results} \label{SectionResults}

In this section, we provide results obtained by implementing the sequential and random quantum circuits defined in Sec.~\ref{SectionTrotter}. First, we show that the sequential quantum circuit with time step $\tau=0.1$ provides a good approximation to the evolution obtained using exact diagonalization. Then, we investigate the evolution of different initial states under random quantum circuits with $\tau=0.1$, and compare to the sequential quantum circuit results. For the states considered here, we observe that random quantum circuits provide a particularly bad approximation to the evolution of non-thermal states, such as states that highly overlap with the quantum many-body scars.

Throughout this section, we use the normalized standard deviation to attest the quality of the results generated by the quantum circuits. The standard deviation we use is defined as:
\begin{align}
    \Delta(Q) = \sqrt{\frac{\sum_{n=0}^N \left[Q(n\tau) - \bar{Q}(n\tau)\right]^2}{\sum_{n=0}^N \bar{Q}(n\tau)^2}}, \label{std}
\end{align}
where $\bar{Q}(n\tau)$ is the reference value for a physical quantity $Q$ at time $t=n\tau$, and $Q(n\tau)$ is the approximate value of $Q$ at time $t=n\tau$. Specifically, to assess the quality of the sequential Trotter evolution, we choose $Q(n\tau)$ and $\bar{Q}(n\tau)$ as quantum expectation values obtained from sequential Trotter evolution and exact evolution. On the other hand, to estimate the effect of randomization of the circuits, $Q(n\tau)$ is the statistical expectation value defined in Eq.~(\ref{statval}) and $\bar{Q}(n\tau)$ is the corresponding value obtained from sequential Trotter evolution. This standard deviation has contributions from all times before $t=N\tau$, hence we are able to discuss the evolution of the standard deviation, which is computed by changing the number of Trotter steps in the sums.

For random quantum circuits, there is an error associated with the standard deviation due to the random choice of lattice sites. In fact, each standard deviation $\Delta(Q)$ comes with a label $k$ associated with the set of random sites used to generate the evolution. If we have a total of $M$ circuit runs available and each standard deviation $\Delta_k(Q)$ is computed using $K$ runs, then the error is given by
\begin{align}
    {\rm Err}(Q) = \sqrt{\frac{\sum_{k=1}^{M/K} \left[\Delta_k(Q) - \bar{\Delta}(Q)\right]^2}{M/K}}, \label{error}
\end{align}
where $\bar{\Delta}(Q)$ is the average of the $M$ standard deviations $\Delta_k(Q)$ available.

\begin{table}
    \centering
    \begin{tabular}{|c|c|c|}
        \hline
         Initial state & $\Delta(\expval{\sigma_{21}^z})$ & $\Delta(\rm{Loschmidt})$ \\
         \hline
         vacuum & $0.056$ & $0.032$ \\
         \hline
         fully-filled & $0.077$ & $0.036$ \\
         \hline
    \end{tabular}
    \caption{Standard deviation of local magnetization and Loschmidt echo, comparing sequential quantum circuit to the exact diagonalization results. We use time step size of $\tau=0.1$ and final time $T=10$.}
    \label{TableSequential}
\end{table}

\begin{figure}
    \includegraphics[scale=0.55]{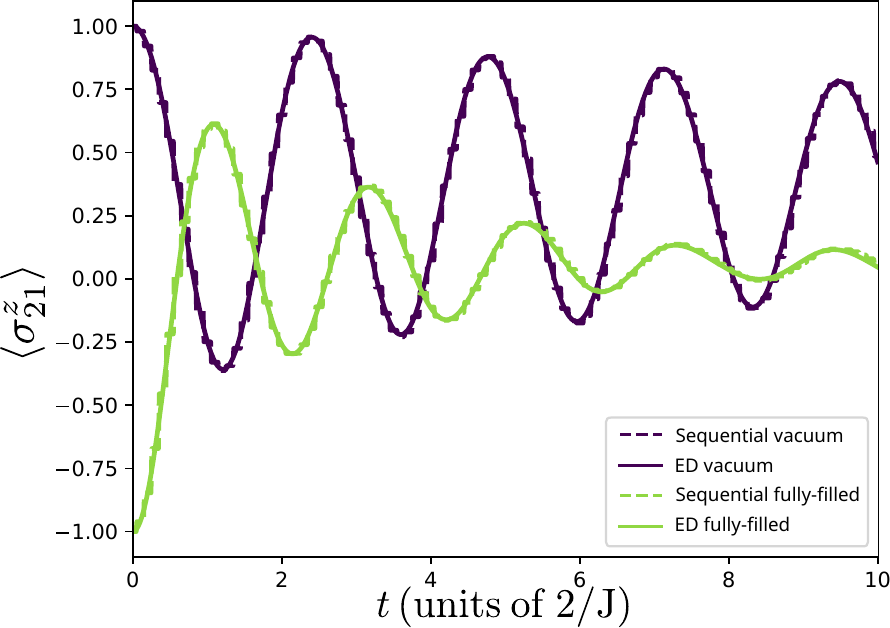}
    \caption{Evolution of local magnetization for the vacuum, purple (dark gray) curves, and the fully-filled state, green (light gray) curves, using exact diagonalization (solid lines) and sequential quantum circuit with time step $\tau=0.1$ (dashed lines). Final time standard deviations are provided in the second column of Table~\ref{TableSequential}.}
    \label{sequentialvsED}
\end{figure}

\paragraph{Sequential quantum circuit.} We simulate the evolution of the local magnetization and the Loschmidt echo of the vacuum and the fully-filled state using sequential quantum circuits with time step size $\tau=0.1$ from $t=0$ to $t=10$. This amounts to $100$ Trotter steps, each consisting of two sets of ten commuting gates, similar to what is represented in Fig.~\ref{circuit}, but in our simulations we use $L=40$ lattice sites. The standard deviations from the sequential quantum circuit to the exact diagonalization results are given in Table~\ref{TableSequential}. In all cases, the standard deviation is smaller than $0.1$, which means that for this time step size the approximation is close to the exact result. Indeed, in Fig.~\ref{sequentialvsED} we plot the evolution of the local magnetization for both states, and visually the sequential and exact curves are quite similar, another indication that the standard deviation $<0.1$ is small.

\begin{figure}
    \includegraphics[scale=0.55]{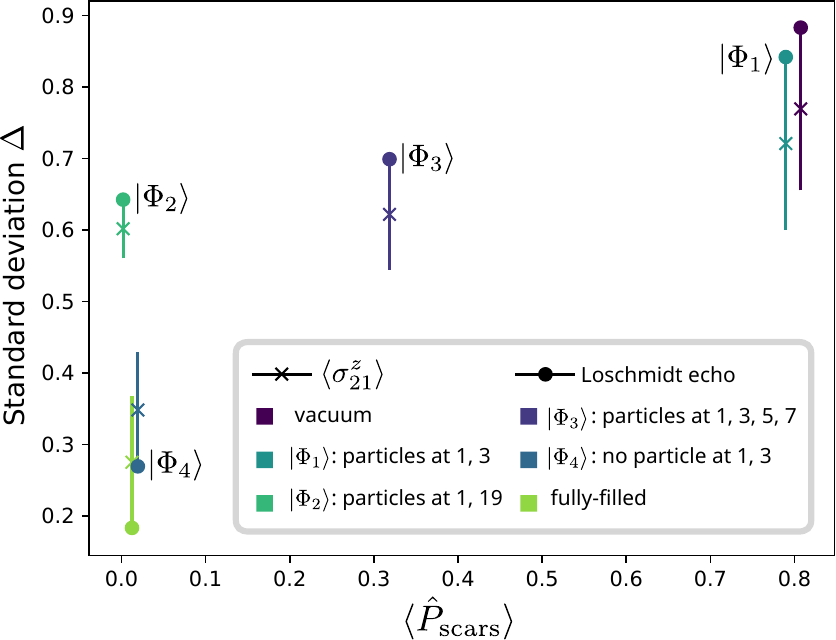}
    \caption{Standard deviation of local magnetization $\expval*{\sigma^z_{21}}$ and Loschmidt echo for different initial states as function of the projection of the initial state in the quantum many-body scars sector. This is the standard deviation of the random quantum circuit evolution with respect to the sequential quantum circuit, with time step size $\tau=0.1$ and duration $T=10$. The lines are error bars calculated using $1000$ circuit runs separated in ten groups of $100$ runs, and the markers are the mean values. }
    \label{stderrorbars}
\end{figure}

\begin{figure}
    \includegraphics[scale=0.55]{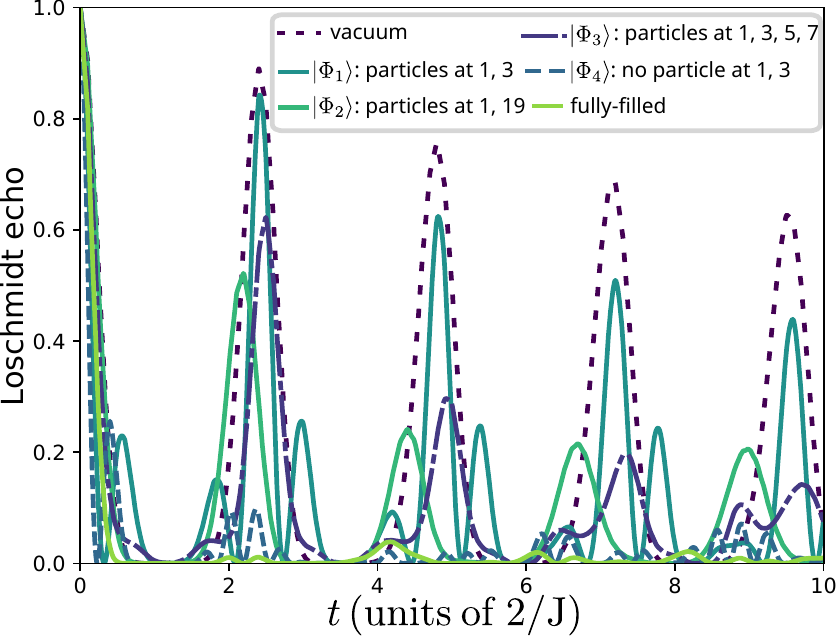}
    \caption{Evolution of the Loschmidt echo for different initial states, computed using sequential quantum circuits with time step size $\tau=0.1$. The Loschmidt echo is the overlap between the evolved state and the initial state.}
    \label{Loschmidtevol}
\end{figure}

\begin{figure}
    \includegraphics[scale=0.55]{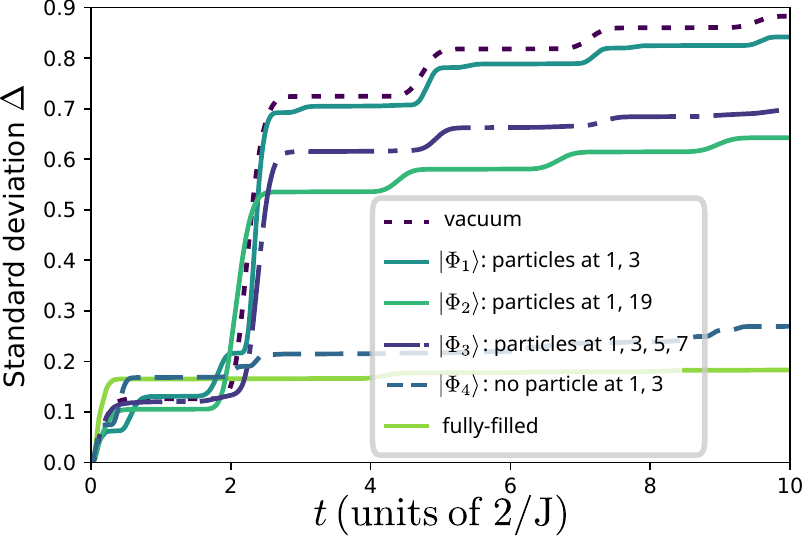}
    \caption{Evolution of the standard deviation of the Loschmidt echo for different initial states. This is the standard deviation of the random quantum circuit evolution with respect to the sequential quantum circuit, with time step size $\tau=0.1$, duration $T=10$. The random quantum circuits were averaged over $K=1000$ runs for each state.}
    \label{stdevol}
\end{figure}

\paragraph{Random quantum circuits.} We maintain time step size $\tau=0.1$, and now use random quantum circuits to simulate the evolution of the local magnetization at the center of the chain, at site $j=21$, and the Loschmidt echo. We simulate $M=1000$ different random quantum circuits for each different initial state, and separate them in ten groups of $K=100$ to compute errors in the standard deviation. The standard deviations for local magnetization and Loschmidt echo are given in Fig.~\ref{stderrorbars}. We show the results as function of the projection in the quantum scar states sector, $\hat{P}_{\rm{scars}}=\sum_s\ket{s}\bra{s}$, where the sum runs over the scar eigenstates of the Hamiltonian. Besides the vacuum and the fully-filled state, we simulate the evolution of other four eigenstates of the two-site translation operator, as in Eq.~\eqref{eqclass}. The fundamental configurations that generate the cyclic families representing the states simulated are: Two particles at sites $j=1$ and $3$ (one pair of particles close to each other), two particles at sites $j=1$ and $19$ (one pair of particles far from each other), four particles at $j=1$, $3$, $5$ and $7$ (two pairs, all particles close to each other), and a configuration full of particles, except for sites $j=1$ and $3$. We label these states as $\ket{\Phi_1}$, $\ket{\Phi_2}$, $\ket{\Phi_3}$ and $\ket{\Phi_4}$, respectively. 

\begin{figure}
    \includegraphics[scale=0.55]{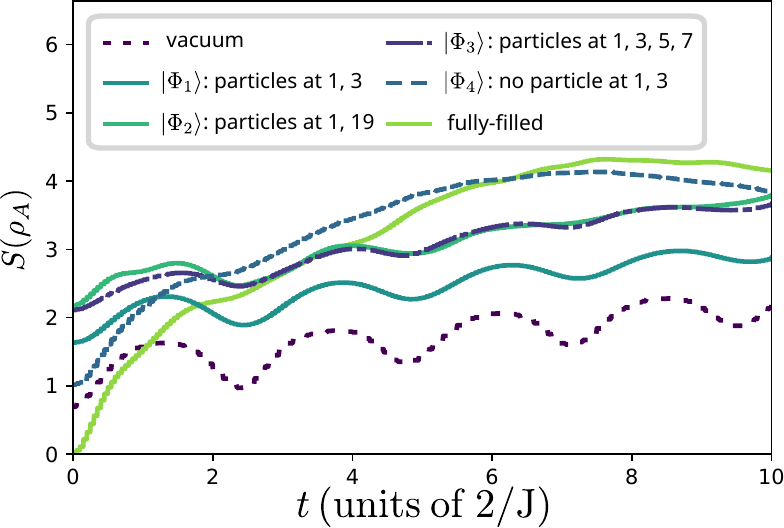}
    \caption{Evolution of the entanglement entropy for different initial states, computed using sequential quantum circuits with time step size $\tau=0.1$. This is the half-chain entanglement entropy: The Hilbert space partition used is $\mathcal{H}_A\otimes\mathcal{H}_B$, where $\mathcal{H}_A$ consists on the first $L/2$ spins, and $\mathcal{H}_B$ on the other $L/2$ spins.}
    \label{EEevol}
\end{figure}

The results in Fig.~\ref{stderrorbars} show that states that highly overlap with the scars, such as the vacuum and $\ket{\Phi_1}$, have high standard deviation of the Loschmidt echo and the local magnetization. If the initial state has low $\expval*{\hat{P}_{\rm{scars}}}$, it does not necessarily mean that the standard deviation is small, which is the case of $\ket{\Phi_2}$ and $\ket{\Phi_3}$. Although these states do not highly overlap with the scars, we observe persistent revivals in the Loschmidt echo in Fig.~\ref{Loschmidtevol}. Such behavior is characteristic of non-thermal states. This is not the case for the fully-filled state or $\ket{\Phi_4}$, they have low overlap with the scars, there no revivals in the Loschmidt echo, and they have low standard deviation. These results reveal that non-thermal states are the ones with higher standard deviation, suggesting that the non-thermal sector of the Hamiltonian is more sensitive to randomness, including the quantum many-body scars. In Fig.~\ref{stdevol}, we plot the evolution of the standard deviation of the Loschmidt echo for the same initial states as in Fig.~\ref{stderrorbars}, but this time all $1000$ runs are averaged out, i.e. there is just one point and no error bars. Before $t=3$ the standard deviation of the Loschmidt echo are all close to each other, but after $t=3$ there is a clear separation between thermal and non-thermal states, and states that highly overlap with the quantum many-body scars have the highest standard deviations. This feature can potentially be used to identify non-thermal states in relatively short simulations. We note that the non-thermal behavior could also be spotted at early times observing the height of the Loschmidt echo peaks between $t=0$ and $t=4$. However, we believe the standard deviation, specially the evolution plot in Fig.~\ref{EEevol}, is a valuable tool as it provides another method to separate thermal and non-thermal states.

\begin{figure}
    \includegraphics[scale=0.55]{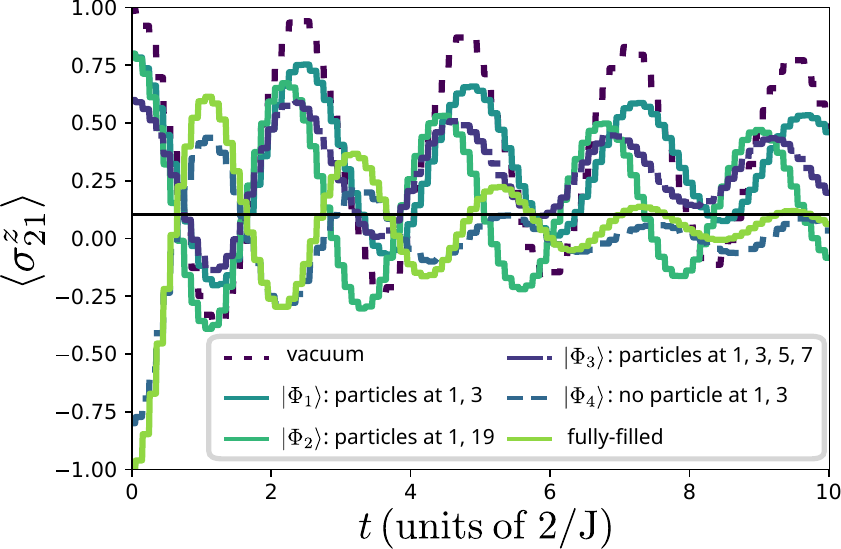}
    \caption{Evolution of the local magnetization $\expval*{\sigma^z_{21}}$ for different initial states, computed using sequential quantum circuits with time step size $\tau=0.1$. The constant curve at $\expval*{\sigma^z_{21}}=0.105$ represents the thermal value of the local magnetization for these initial states.}
    \label{Zmidevol}
\end{figure}

\paragraph{Thermal and non-thermal states.} In the last paragraph, we discussed how the standard deviation and the peaks in the Loschmidt echo can be used to separate thermal from non-thermal states. In this paragraph, we use the entanglement entropy and the local magnetization as further proof to categorize the states. In Fig.~\ref{EEevol}, we used sequential quantum circuits of time step size $\tau=0.1$ to simulate the evolution of the half-chain entanglement entropy for the different initial states introduced in the previous paragraphs: Vacuum state, fully-filled state, $\ket{\Phi_1}$, $\ket{\Phi_2}$, $\ket{\Phi_3}$ and $\ket{\Phi_4}$. Comparing the entanglement entropy values at $t=0$ and $t=5$, we observe that the fully-filled state and the state $\ket{\Phi_4}$ are the ones that increased faster in this period. On the other hand, in this same period the vacuum, $\ket{\Phi_1}$, $\ket{\Phi_2}$ and $\ket{\Phi_3}$ had slower growth of the entanglement entropy. We expect thermal states to have fast growth of entanglement entropy, hence Fig.~\ref{EEevol} is an evidence that the vacuum state, $\ket{\Phi_1}$, $\ket{\Phi_2}$ and $\ket{\Phi_3}$ are non-thermal states, while the fully-filled state and $\ket{\Phi_4}$ are thermal. We also used sequential quantum circuits of time step size $\tau=0.1$ to simulate the evolution of the local magnetization in Fig.~\ref{Zmidevol}. The vacuum, $\ket{\Phi_1}$, $\ket{\Phi_2}$ and $\ket{\Phi_3}$ highly oscillate in the period from $t=0$ to $t=10$, while the fully-filled state and $\ket{\Phi_4}$ are the states that considerably decrease their oscillation amplitude to approach their thermal value of $\expval*{\sigma^z_{21}}=0.105$ in this time scale.

\section{Discussion} \label{SectionConclusion}

The discovery of quantum many-body scar states is one of the exciting results in the field of quantum simulations of lattice gauge theories. In a previous work \cite{Andrade2022}, we proposed a strategy for quantum simulation of the Schwinger model using three-body gates engineered by Raman beams and trapped ions. In this work, we use the mapping from Ref.~\cite{Surace2020} between the Schwinger model and the PXP model to identify the quantum many-body scars in the Schwinger model. Then we compare the behavior of thermal and non-thermal states under evolution generated by sequential and randomized quantum circuits. 

The sequential and random quantum circuits used are defined in Section~\ref{SectionTrotter}, and the quantum many-body scars are described in Section~\ref{SectionScars} and in Fig.~\ref{scars_qlm}, where we show how they highly violate the Eigenstate Thermalization Hypothesis. In Section~\ref{SectionResults}, we explored the dynamics of physical states using Trotterization with quantum circuits consisting of three-body gates as engineered in our previous work \cite{Andrade2022}. We used the normalized standard deviation defined in Eq.~\eqref{std} to evaluate the quality of different quantum circuits. For a quantum simulation of total time $T=10$, sequential quantum circuits with $N=100$ Trotter steps, or $\tau=0.1$, provide a good approximation to the exact diagonalization results, with standard deviations smaller than $0.1$, as described in Table~\ref{TableSequential}. Each Trotter step consists of $20$ three-body gates, which is the number of terms in the Schwinger QLM Hamiltonian of Eq.~\eqref{QLMHamiltonian} with $L=40$ sites. In Fig.~\ref{sequentialvsED} we compare the evolution of the local magnetization generated by a sequential circuit with $\tau=0.1$ to the exact evolution. This direct comparison for the vacuum and the fully-filled state corroborate that standard deviations of $<0.1$ are quite small. 

The standard deviations are larger when using random quantum circuits to generate the evolution, and in this case we have to consider errors, defined in Eq.~\eqref{error}, coming from the different choices of random sites for each random quantum circuit. We generated $1000$ different random quantum circuit to perform simulations for six initial states. In each case, we used the random quantum circuits to evolve the Loschmidt echo and the local magnetization.  

In Fig.~\ref{stderrorbars} we have the standard deviation at time $t=10$ as function of the projection of the initial state in the scars sector. We observe a similar trend for both Loschmidt echo and local magnetization: States that highly overlap with the scars have high standard deviation of the random quantum circuits with respect to the sequential one, which is the case of the vacuum and the state $\ket{\Phi_1}$. On the other hand, the fully-filled state and states $\ket{\Phi_2}$, $\ket{\Phi_3}$ and $\ket{\Phi_4}$ have lower projection in the scars sector, with $\ket{\Phi_2}$ and $\ket{\Phi_3}$ having the highest standard deviations. In Fig.~\ref{Loschmidtevol} we show the evolution of the Loschmidt echo for all states simulated, the evolution here is generated by a sequential circuit with $\tau=0.1$. The revivals in the Loschmidt echo only happen for states that have of high standard deviation in Fig.~\ref{stderrorbars}, which is a suggestion that the non-thermal states are more sensitive to randomization. The non-thermal behavior of the states $\ket{\Phi_1}$, $\ket{\Phi_2}$, $\ket{\Phi_3}$ and the vacuum is evident from the presence of revivals in the Loschmidt echo, from the slower growth of entanglement as shown in Fig.~\ref{EEevol}, and the persistent oscillations away from the thermal value in the local magnetization in Fig.~\ref{Zmidevol}. The results obtained from the initial states simulated encourage us to believe that the non-thermal sectors of the Schwinger QLM Hamiltonian, including the quantum many-body scars, are destroyed by the presence of random gates as the gates now promote mixing of the non-thermal states with states in the thermal sector.

Upon trying to construct an effective Hamiltonian to capture the behavior of the scar states under evolution by a random circuit, we observe that the random gates do not respect the 2-site translation symmetry of the original Hamiltonian, hence we need to consider a Hilbert space without the equivalence classes of Eq.~\eqref{eqclass}. This increases the Hilbert space size by a factor of $\sim L/2$, and to identify the new scars we need to perform several exact diagonalizations, which becomes computationally costly even if we consider only a few random quantum circuit runs. Another direction for a continuation of this work would be exploring different classes of quantum scars, which have been identified for the PXP model \cite{Szoldra2022}. Moreover, by including local measurements into the random circuits, one might investigate entanglement transitions  \cite{Skinner2019}, at a critical measurement rate which we expect to differ significantly between scar states and generic states.

\begin{acknowledgements}

BA acknowledges funding from the European Union's Horizon 2020 research and innovation programme under the Marie Sk{\l}odowska-Curie grant agreement No. 847517.
This work has been financially supported by the Ministry of Economic Affairs and Digital Transformation of the Spanish Government through the QUANTUM ENIA project call – Quantum Spain project, and by the European Union through the Recovery, Transformation and Resilience Plan – NextGenerationEU within the framework of the Digital Spain 2026 Agenda.
ICFO group acknowledges support from:
ERC AdG NOQIA; MCIN/AEI (PGC2018-0910.13039/501100011033, CEX2019-000910-S/10.13039/501100011033, Plan National FIDEUA PID2019-106901GB-I00, Plan National STAMEENA PID2022-139099NB-I00 project funded by MCIN/AEI/10.13039/501100011033 and by the “European Union NextGenerationEU/PRTR" (PRTR-C17.I1), FPI); QUANTERA MAQS PCI2019-111828-2); QUANTERA DYNAMITE PCI2022-132919 (QuantERA II Programme co-funded by European Union’s Horizon 2020 program under Grant Agreement No 101017733), Ministry of Economic Affairs and Digital Transformation of the Spanish Government through the QUANTUM ENIA project call – Quantum Spain project, and by the European Union through the Recovery, Transformation, and Resilience Plan – NextGenerationEU within the framework of the Digital Spain 2026 Agenda; Fundació Cellex; Fundació Mir-Puig; Generalitat de Catalunya (European Social Fund FEDER and CERCA program, AGAUR Grant No. 2021 SGR 01452, QuantumCAT \ U16-011424, co-funded by ERDF Operational Program of Catalonia 2014-2020); Barcelona Supercomputing Center MareNostrum (FI-2023-1-0013); EU Quantum Flagship (PASQuanS2.1, 101113690); EU Horizon 2020 FET-OPEN OPTOlogic (Grant No 899794); EU Horizon Europe Program (Grant Agreement 101080086 — NeQST), ICFO Internal “QuantumGaudi” project; European Union’s Horizon 2020 program under the Marie Sklodowska-Curie grant agreement No 847648;  “La Caixa” Junior Leaders fellowships, La Caixa” Foundation (ID 100010434): CF/BQ/PR23/11980043. Views and opinions expressed are, however, those of the author(s) only and do not necessarily reflect those of the European Union, European Commission, European Climate, Infrastructure and Environment Executive Agency (CINEA), or any other granting authority.  Neither the European Union nor any granting authority can be held responsible for them.
T.G. acknowledges funding by Gipuzkoa Provincial Council (QUAN-000021-01), by the Department of Education of the Basque Government through the IKUR strategy and through the project PIBA\_2023\_1\_0021 (TENINT), by the Agencia Estatal de Investigación (AEI) through Proyectos de Generación de Conocimiento PID2022-142308NA-I00 (EXQUSMI), and that the work has been produced with the support of a 2023 Leonardo Grant for Researchers in Physics, BBVA Foundation. The BBVA Foundation is not responsible for the opinions, comments and contents included in the project and/or the results derived therefrom, which are the total and absolute responsibility of the authors. 
U.B. is also grateful for the financial support of the IBM Quantum Researcher Program.
R.W.C acknowledges support from the Polish National Science Centre (NCN) under the Maestro Grant No. DEC-2019/34/A/ST2/00081.  

The Authors would like to thank Piotr Sierant for his comments and suggestions on the first version of this manuscript.

\end{acknowledgements}

\newpage 

\nocite{*}
\bibliography{bibliography.bib}

\end{document}